\documentclass[showpacs,amsmath,
twocolumn,
aps,prl]{revtex4}
\usepackage{graphicx}
\usepackage{dcolumn}
\usepackage[colorlinks=true, pdfstartview=FitV, linkcolor=blue, 
            citecolor=blue, urlcolor=blue]{hyperref}

\begin{document}

\title{Nuclear magnetic resonance force microscopy with a microwire rf
  source} \author{M.  Poggio$^{1,2}$, C. L.  Degen$^1$, C. T.
  Rettner$^1$, H.  J.  Mamin$^1$, and D.  Rugar$^1$}
\affiliation{$^1$IBM Research Division,
  Almaden Research Center, 650 Harry Rd., San Jose CA, 95120 \\
  $^2$Center for Probing the Nanoscale, Stanford University, 476
  Lomita Hall, Stanford CA, 94305} \date{\today}

\begin{abstract}
  
  We use a 1.0-$\mu$m-wide patterned Cu wire with an integrated
  nanomagnetic tip to measure the statistical nuclear polarization of
  $^{19}$F in CaF$_2$ by magnetic resonance force microscopy (MRFM).
  With less than 350 $\mu$W of dissipated power, we achieve rf
  magnetic fields over 4 mT at 115 MHz for a sample positioned within
  100 nm of the ``microwire'' rf source.  A 200-nm-diameter FeCo tip
  integrated onto the wire produces field gradients greater than
  $10^5$ T/m at the same position.  The large rf fields from the
  broadband microwire enable long rotating-frame spin lifetimes of up
  to 15 s at 4 K.
  
\end{abstract}

\pacs{85.85+j, 85.35.-p, 81.16.-c, 84.40.Az}

\maketitle

The proposal of magnetic resonance force microscopy (MRFM)
\cite{Sidles:1992} and its subsequent realization \cite{Rugar:1992}
combine the physics of magnetic resonance imaging with the techniques
of scanning probe microscopy.  Recently this marriage has led to the
demonstration of nuclear spin imaging with a spatial resolution of 90
nm \cite{Mamin:2007}.  In order to eventually image on the scale of
single nuclear spins, the force sensitivity of the measurement must be
improved by roughly 3 orders of magnitude.  Such an improvement will
only be achieved if the dimensions of some key components are scaled
down to more closely match the nanometer and sub-nanometer length
scales of single spin physics.  The most recent advances in
sensitivity were the result of an increase in the magnetic field
gradient provided by the scanning magnetic tip.  Here, we discuss MRFM
measurements done with a ``microwire'' rf source and an integrated
nanomagnetic tip meant to further scale down the measurement
apparatus.  The reduced heat dissipation of the compact new geometry
allows us to simultaneously access temperatures under 300 mK and rf
magnetic field amplitudes above 4 mT --- fields large enough to
produce remarkably long rotating-frame nuclear spin lifetimes.

We measure the statistical polarization \cite{Mamin:2003} of $^{19}$F
spins in a CaF$_2$ crystal using a technique known as adiabatic rapid
passage \cite{Slichter:1990}.  In a fixed magnetic field $\bf{B_0}$,
we sweep the frequency $\nu_{\text{rf}}$ of a transverse rf magnetic
field $\bf{B_1}$ through the nuclear resonance condition,
$\nu_{\text{rf}} = \frac{\gamma}{2 \pi} B_0$, where $\gamma$ is the
gyromagnetic ratio of $^{19}$F.  If done slowly enough, i.e. if the
adiabatic condition $\partial \nu_{\text{rf}} / \partial t \ll
\frac{\gamma^2}{2 \pi} B_1^2$ is met, then the sweep induces nuclear
spin inversions along the $\bf{B_0}$ direction.  We then detect these
inversions using a magnetic tip and an ultrasensitive cantilever as a
force detector.

\begin{figure}[t]\includegraphics{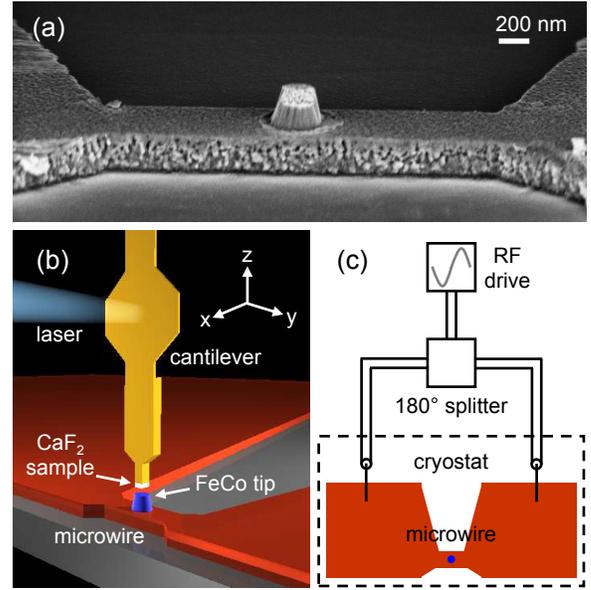}
\caption{\label{fig1}
  (a) Scanning electron micrograph of the Cu microwire with integrated
  FeCo tip.  (b) Representation of the experimental apparatus at the
  bottom of the cryostat (the relative scale of the components has
  been slightly altered).  $\bf{B_0}$, the cantilever shaft, and the
  axis of the magnetic tip are aligned along $\hat{z}$.  Current flows
  in the wire along $\hat{y}$, while at the position of the sample,
  the lever displacement and $\bf{B_1}$ are directed along $\hat{x}$.
  (c) Schematic diagram of the electrical connections to the
  microwire.  }\end{figure}

\begin{figure}[t]\includegraphics{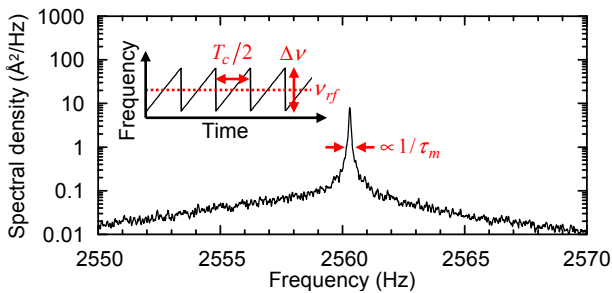}
\caption{\label{fig2}
  The power spectral density of cantilever displacement during
  adiabatic rapid passage of a statistical polarization of $^{19}$F
  spins.  The inset represents the frequency sweep of $\bf{B_1}$ that
  induces the passages.  }\end{figure}

The force detection apparatus, shown in Fig.~\ref{fig1}(b), uses a
sample-on-cantilever configuration.  The single crystal Si cantilever
is 120-$\mu$m long, 3-$\mu$m wide, and 0.1-$\mu$m thick and includes a
15-$\mu$m long, 2-$\mu$m thick mass on its end \cite{Chui:2003}.  The
cantilever's mass-loaded geometry suppresses the motion of flexural
modes above the fundamental frequency
\cite{Mozyrsky:2003,Berman:2003}.  A $\sim 50$-$\mu$m$^3$ particle of
CaF$_2$ crystal glued to the end of the lever serves as the sample.  A
thin layer of Si/Au (10/30 nm), with Si as an adhesion layer, is
evaporated onto the end of the sample to screen electrostatic fields.
At $T = 4.2$ K the sample-loaded cantilever has a resonant frequency
$\nu_c = 2.6$ kHz and an intrinsic quality factor $Q_0 = 44,000$.  The
oscillator's spring constant is determined to be $k = 86$ $\mu$N/m
through measurements of its thermal noise spectrum at several
different base temperatures.  The cantilever is mounted in a vacuum
chamber (pressure $< 1 \times 10^{-6}$ torr) at the bottom of a
dilution refrigerator, which is isolated from environmental
vibrations.  The motion of the lever is detected using laser light
focused onto a 10-$\mu$m wide paddle near the mass-loaded end and
reflected back into an optical fiber interferometer \cite{Rugar:1989}.
100 nW of light are incident on the paddle from a temperature-tuned
1550-nm distributed feedback laser diode \cite{Bruland:1999}.  We damp
the cantilever using feedback to a quality factor of $Q = 250$ in
order to increase the bandwidth of our force detection without
sacrificing force sensitivity \cite{Garbini:1996}.

The key component of this experiment is the microwire rf source, which
efficiently produces a strong rf magnetic field $\bf{B_1}$ for nuclear
magnetic resonance.  The Cu wire is 2.6-$\mu$m long, 1.0-$\mu$m wide,
and 0.2-$\mu$m thick and is patterned atop a Si substrate, as shown in
Fig.~\ref{fig1}(a).  The microwire bridges the gap between two
1-mm$^2$ pads and has a resistance of 0.35 $\Omega$ at $ T = 4.2$ K.
In the middle of the microwire structure, deposited on its surface, is
a 250-nm tall, 200-nm wide FeCo tip, in the shape of a truncated cone.
This nanomagnetic tip provides the spatial magnetic field gradient
required by the MRFM measurement.

We fabricate the microwire using lift-off, then we place the magnetic
tip on the wire through a novel stencil-based process.  First, a
450-nm layer of IBM KRS photoresist is patterned using electron-beam
lithography on a pre-scribed wafer to define the copper wire.  A
Cr/Cu/Au (5/200/5 nm) film is then deposited via thermal evaporation
and lifted off in hot solvent with ultrasonic agitation.  To form the
magnetic tip, a 500-nm thick film of polyimide is then spun onto the
wafer and coated with a thin layer of evaporated Ti.  A single hole is
written above the wire, again with electron beam lithography, in a
resist covering the Ti.  The Ti is then etched with a CF$_4$ plasma.
Next we etch the polyimide with an O$_2$ reactive ion etch through
this hole to form a cavity.  The resulting undercut Ti/polyimide
bilayer structure forms a stencil mask over the wire onto which the
magnetic film can be evaporated.  This film consists of
Ti/Co$_{30}$Fe$_{70}$/Au (15/200/15 nm) deposited by e-beam
evaporation.  All the Au capping layers are meant to provide
protection against oxidation.  After final lift-off, the wafer is
cleaved along the pre-existing scribe in order to place the wire less
than 50 $\mu$m from the edge of the chip.  The structure's proximity
to the edge ensures a clear optical path from the fiber interferometer
to the cantilever paddle.

\begin{figure}[t]\includegraphics{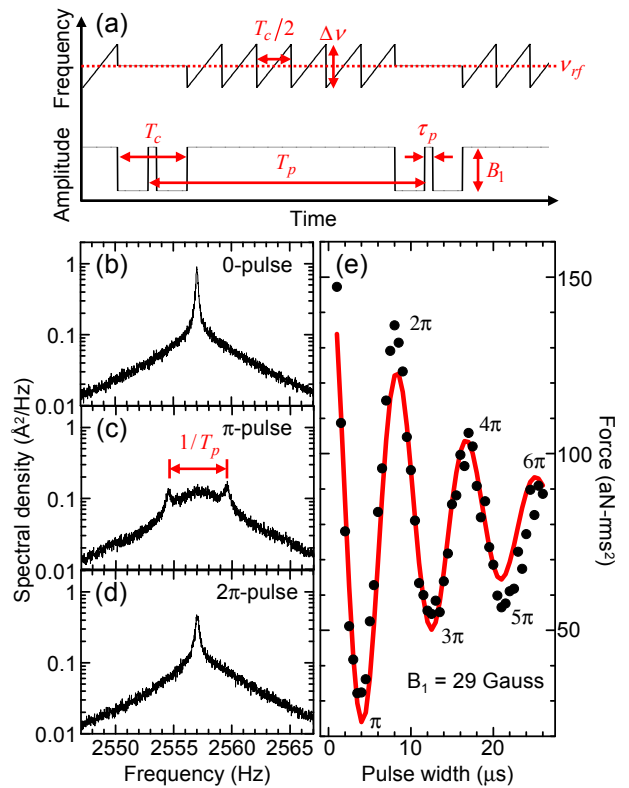}
\caption{\label{fig3}
  (a) The rf pulse protocol for the spin nutation experiment.  The
  resulting power spectral density of cantilever displacement with
  pulse widths $\tau_p$ equivalent to (b) 0, (c) $\pi$, and (d) $2
  \pi$ radians of nutation.  (e) The corresponding force signal is
  measured through a narrow band lock-in amplifier and plotted in
  points as a function of pulse width $\tau_p$.  A $B_1$ amplitude can
  be extracted from a decaying cosinusoidal fit of the Rabi
  oscillations shown in red.  }\end{figure}

The overall geometry of the MRFM apparatus is shown in
Fig.~\ref{fig1}(b).  During measurement, the sample at the end of the
cantilever is situated less than 100 nm directly above the
nanomagnetic tip.  At such a small spacing, the magnetic tip provides
fixed spatial field gradients in excess of $10^5$ T/m.  Less than 20
mA passing through the microwire (current density $\sim 10^7$
A/cm$^2$) produce rf $B_1$ fields larger than 4 mT (rotating field) at
the position of the sample.  Under these conditions, the heat
dissipated by the wire is under 350 $\mu$W allowing the dilution
refrigerator to reach temperatures below 300 mK.  The rf current
required to produce $B_1$ flows through the microwire from the pads on
either side of it.  These pads are each connected through short Cu
leads to the center conductors of semi-rigid coaxial lines leading to
the top of the cryostat.  The two lines are differentially driven by a
180$^\circ$ splitter, which is in turn connected to the rf drive
signal as shown in Fig.~\ref{fig1}(c).  By matching the attenuation
and delay caused by each coaxial line, this differential driving
scheme results in a voltage node and a current anti-node at the
microwire.  In this way, we maximize the rf magnetic field produced
around the microwire while minimizing unwanted electric fields, which
could cause spurious excitation of our cantilever.  Note that since
the wire is not frequency-specific it has the flexibility of being a
broadband source of rf magnetic field.  In the present experiment, the
upper frequency limit is about 200 MHz due to the inductance of the
wire leads.

In typical MRFM measurements of the $^{19}$F spin polarization by
adiabatic rapid passage, we drive the microwire with the
frequency-sweep waveform represented in the inset to Fig.~\ref{fig2}.
In these experiments we use a center frequency $\nu_{\text{rf}} =
114.7$ MHz and a peak-to-peak frequency deviation $\Delta \nu = 1.4$
MHz.  A superconducting magnet provides the resonant field for
$^{19}$F of $B_0 \simeq 2.9$ T.  By generating an rf magnetic field
whose frequency is swept through the $^{19}$F resonance twice every
cantilever period $T_c$, we drive longitudinal nuclear spin flips in
the sample at the lever's resonance frequency.  Since the sample is
mounted on the end of the cantilever, in the presence of a large
enough magnetic field gradient, the spin flips produce a force that
drives the lever.  By measuring the amplitude of the cantilever's
oscillation on resonance we determine the longitudinal component of
net spin polarization.  In our case, this polarization is due to the
naturally occurring $\sqrt{N}$ statistical component.  The cantilever
displacement induced by adiabatic passages is shown in the vibrational
spectrum plotted in Fig.~\ref{fig2}.  The narrow band spin signal,
whose spectral width is inversely proportional to the rotating-frame
spin lifetime $\tau_m$, sits atop a much broader peak generated by the
lever's natural thermal vibrations.

To measure the magnitude of $B_1$, which the microwire produces in the
sample, a different rf protocol, shown schematically in
Fig.~\ref{fig3}(a), is used.  We sweep the frequency $\nu_{\text{rf}}$
of $B_1$ through resonance twice per cantilever period, as discussed
previously, and intersperse the frequency sweeps with occasional
resonant pulses of variable width $\tau_p$ , spaced by $T_p$.  Here we
use $T_p = 196$ ms or $502 \times T_c$.  Given a fixed amplitude
$B_1$, as we increase $\tau_p$, the resonant pulses induce spins to
nutate with an increasing angle.  If the pulse spacing is much less
than the rotating-frame spin lifetime, $T_p \ll \tau_m$, nutations
produced by the pulses modulate the force signal generated by the
adiabatic passages.  When $\tau_p = \pi / (\gamma B_1)$, i.e.\ the
pulse width and amplitude is equivalent to $\pi$ radians of nutation,
each pulse reverses the sign of the force signal.  This modulation
results in sidebands appearing in the frequency spectrum of the
cantilever displacement signal spaced by $1/(2 T_p)$ from the lever
resonance at $\nu_c$.  The signal power formerly in the central peak
shifts to the sidebands and back again depending on the rf pulse width
$\tau_p$, as shown in Fig.~\ref{fig3}(b)-(d).  By feeding the
displacement signal into a narrow band lock-in amplifier referenced to
$\nu_c$, we measure the power at $\nu_c$ as a function of $\tau_p$ and
observe the Rabi oscillations plotted in Fig.~\ref{fig3}(e).  From the
period of these oscillations we extract the pulse width required for
$2 \pi$ radians of nutation, $\tau_{2 \pi}$.  Using the relation $B_1
= 2 \pi / (\gamma \tau_{2 \pi})$, we determine the magnitude of $B_1$
produced by the microwire at the sample position.  Aquiring data
similar to those shown in Fig.~\ref{fig3}(e) taken at different rf
drive amplitudes, we calibrate the magnitude of $B_1$ at the sample
position corresponding to a given rf drive.

\begin{figure}[b]\includegraphics{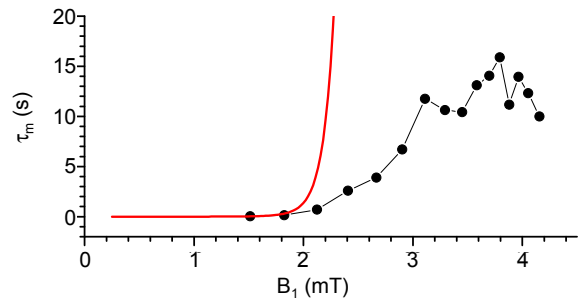}
\caption{\label{fig4}
  Plot of the $\tau_m$ as a function of $B_1$ in the sample at $T =
  4.2$ K.  Red line corresponds to the limit set on $\tau_m$ by the
  adiabatic condition and is calculated using numerical integration of
  the Bloch equations for repetitive linear frequency sweeps of
  $\nu_{\text{rf}}$ through resonance \cite{Baum:1985}. }\end{figure}

In order to investigate the dependence of $\tau_m$ for $^{19}$F nuclei
in CaF$_2$ as a function of increasing rf magnetic field $B_1$, we
employ the adiabatic sweep waveform without the interspersed pulses,
shown in the inset to Fig.~\ref{fig2}.  As plotted in Fig.~\ref{fig4},
$\tau_m$ strongly increases with increasing $B_1$ amplitude up to a
saturation around $\tau_m = 15$ s at $B_1 = 3$ mT.  Previous
low-temperature nuclear MRFM experiments were done with $B_1 < 2$ mT
due to the large heat dissipation caused by the larger than
100-$\mu$m-diameter coils used as rf sources \cite{Mamin:2007}.  These
conditions resulted in $\tau_m < 500$ ms.  In this low $B_1$ regime,
rotating-frame lifetimes are often very short due to either violation
of the adiabatic condition \cite{Miller:2005}, spin-spin interactions,
or spin relaxation caused by the thermal vibration of higher order
cantilever modes in the strong field gradients provided by the
nanomagnetic tip \cite{Mozyrsky:2003,Berman:2003}.

In addition to the newly accessible regime of long $\tau_m$, the small
amount of heat dissipated by the microwire --- even for large $B_1$
amplitudes --- is the fundamental advance presented here.  Previous
nuclear MRFM experiments were done with a hand-wound coil larger than
200 $\mu$m in diameter as an rf source; the coil produced less than 2
mT at the sample with more than 200 mW of dissipated heat.  In
contrast, since our microwire can be less than 100 nm from the sample,
it produces more than 4 mT with less than 350 $\mu$W of dissipated
heat.

\begin{acknowledgments}
  We thank M. Hart and M. Farinelli for assistance with magnetic tip
  fabrication.  We acknowledge support from the DARPA QuIST program
  administered through the Army Research Office, the NSF-funded Center
  for Probing the Nanoscale (CPN) at Stanford University, and the
  Swiss National Science Foundation.
\end{acknowledgments}


\begin{thebibliography}{21}
\bibitem{Sidles:1992} J. A. Sidles,
  \href{http://dx.doi.org/10.1103/PhysRevLett.68.1124}{\textit{Phys.
      Rev. Lett.}  \textbf{68}, 1124 (1992)}.
\bibitem{Rugar:1992} D. Rugar, C. S. Yannoni, and J. A. Sidles,
  \href{http://dx.doi.org/10.1038/360563a0}{\textit{Nature}
    \textbf{360}, 563 (1992)}.
\bibitem{Mamin:2007} H. J. Mamin, M. Poggio, C. L. Degen, and D.
  Rugar,
  \href{http://dx.doi.org/10.1038/nnano.2007.105}{\textit{Nature
      Nanotechnology} \textbf{2}, 301 (2007)};
  \href{http://arxiv.org/abs/cond-mat/0702664}{cond-mat/0702664}.
\bibitem{Mamin:2003} H. J. Mamin, R. Budakian, B. W. Chui, and D.
  Rugar,
  \href{http://dx.doi.org/10.1103/PhysRevLett.91.207604}{\textit{Phys.
      Rev. Lett.} \textbf{91}, 207604 (2003)}.
\bibitem{Slichter:1990} C. P. Slichter, \textit{Principles of Magnetic
    Resonance} (Springer-Verlag, New York, 1990), 3rd ed., Chaps. 2
  and 6.
\bibitem{Chui:2003} B. W. Chui \textit{et al.}, \textit{Technical
    Digest of the 12th International Conference on Solid-State Sensors
    and Actuators (Transducers '03)}, (IEEE Boston, MA, 2003), p.
  1120.
\bibitem{Mozyrsky:2003} D. Mozyrsky, I. Martin, D. Pelekhov, and P. C.
  Hammel, \href{http://dx.doi.org/10.1063/1.121046}{\textit{Appl.
      Phys. Lett.} \textbf{82}, 1278 (2002)}.
\bibitem{Berman:2003} G. P. Berman, V. N. Gorshkov, D. Rugar, and V.
  I. Tsifrinovich,
  \href{http://dx.doi.org/10.1103/PhysRevB.68.094402}{\textit{Phys.
      Rev. B} \textbf{68}, 094402 (2003)}.
\bibitem{Rugar:1989} D. Rugar, H. J. Mamin, and P. Guethner,
  \href{http://dx.doi.org/10.1063/1.101987}{\textit{Appl.  Phys.
      Lett.} \textbf{55}, 2588 (1989)}.
\bibitem{Bruland:1999} K. J. Bruland \textit{et al.},
  \href{http://dx.doi.org/10.1063/1.1149947}{\textit{Rev. Sci. Inst.}
    \textbf{70}, 3542 (1999)}.
\bibitem{Garbini:1996} J. L. Garbini, K. J. Bruland, W. M. Dougherty,
  and J. A. Sidles,
  \href{http://dx.doi.org/10.1063/1.363085}{\textit{J. Appl. Phys.}
    \textbf{80}, 1951 (1996)}; K. J. Bruland, J. L. Garbini, W. M.
  Dougherty, and J. A. Sidles,
  \href{http://dx.doi.org/10.1063/1.363086}{\textit{J. Appl. Phys.}
    \textbf{80}, 1959 (1996)}.
\bibitem{Miller:2005} C. W. Miller and J. T. Markert,
  \href{http://dx.doi.org/10.1103/PhysRevB.72.224402}{\textit{Phys.
      Rev. B} \textbf{72}, 224402 (2005)}.
\bibitem{Baum:1985} J. Baum, R. Tycko, and A. Pines,
  \href{http://dx.doi.org/10.1103/PhysRevA.32.3435}{\textit{Phys. Rev.
      A} \textbf{32}, 3435 (1985)}.

\end{thebibliography}
\end{document}